\documentclass[letterpaper,11pt]{article} 
\usepackage[letterpaper, margin=1in]{geometry}
\usepackage{graphicx}
\usepackage{cite}
\usepackage{amsmath}
\usepackage{amssymb}
\usepackage{url}
\usepackage{dblfloatfix}
\usepackage[bookmarks=false]{hyperref}
\usepackage{color}
\graphicspath{{Figures/}}

\begin{document}

\title{\LARGE \bf A Metric for Evaluating and Comparing Closed-Loop Deep Brain Stimulation Algorithms}

\author{Jeffrey Herron$^{1,5}$, Anca Velisar$^{2}$, Mahsa Malekmohammadi$^{2}$, \\
Helen Bronte-Stewart$^{2,3}$, and Howard Jay Chizeck$^{1,4,5}$\\ \\
$^{1}$Department of Electrical Engineering, University of Washington, Seattle, WA\\
$^{2}$Department of Neurology and Neurological Sciences, Stanford University, Stanford, CA\\
$^{3}$Department of Neurosurgery, Stanford University, Stanford, CA\\
$^{4}$Department of Bioengineering, University of Washington, Seattle, WA\\
$^{5}$NSF Engineering Research Center for Sensorimotor Neural Engineering
}
\maketitle

\begin{abstract}
\textit{Objective.} Closed-loop deep brain stimulation (DBS) may improve current clinical DBS treatment for neurological movement disorders, but control algorithms may perform differently across patients. New metrics are needed for comparing and evaluating closed-loop algorithm performance that address the specific needs of closed-loop neuromodulation controllers. \textit{Approach.} A metric is proposed for system performance that includes normalized terms that can be used to compare algorithm performance for a patient. This metric was evaluated using two closed-loop control algorithms that were tested in patients with Parkinson's Disease (PD) who experience rest tremor. \textit{Main Results.} The metric's resulting balance between tremor treatment and power savings varied on a per patient and algorithm basis. This was expected given how each trial resulted in a variable reduction in stimulation power at the cost of additional tremor for the patient when compared to open-loop stimulation. \textit{Significance.} The proposed metric will aid in clinical evaluation of new algorithms and provide a benchmark for future system designers. This will be important given the growing potential applications of dynamically adjusted neural stimulation.  ClinicalTrials.gov Identifier: NCT02384421.
\end{abstract}

{\bfseries{\textit{\small Keywords}}: {\it{Deep Brain Stimulation, Parkinson's Disease, Closed-Loop Systems, \\Wearable Sensing}}}

\normalsize
\normalfont

\section{Introduction}\label{sec:Introdcution}
DBS has proven to be a safe and effective method of treating Parkinson's disease (PD)\cite{deuschl2006PDrandomized}. However, the selection of stimulation parameters during device programming is empirical and does not always result in the most effective treatment \cite{kuncel2006clinical,mcintyre2015engineering}. Additionally, the stimulation of brain tissue can cause a variety of unintended side effects \cite{kuncel2006clinical}. Existing DBS systems do not take into account the variable nature of symptoms that result from neurological movement disorders \cite{hebb2014creating_fb,sun2014closed}. In addition, these open-loop systems may inefficiently use battery and unnecessarily expose patients to side-effects due to the fact that they stimulate deep brain structures at a constant level and consistent rate.

One solution to address this problem is to create a feedback loop where DBS parameters are adjusted based on the severity of symptoms that a patient experiences in `real time'. The few fully integrated closed-loop systems tested with human patients have so far shown promise for improving DBS treatment. Sensed beta-band information has been used to turn on and off stimulation in eight patients with recordings that were taken from externalized leads in the perioperative period  \cite{little2013adaptive}. In a different study, four Essential Tremor patients, who exhibited severe intention tremor, were tested with a closed-loop DBS system that determined and triggered stimulation based on voluntary muscle activity \cite{yamamoto2013_on_demand_dbs}.

One issue that we have observed while developing closed-loop control algorithms for PD rest tremor\cite{mahsa2015feasibility}, is that any given algorithm can produce dramatically different results across patients or even for the same patient on different limbs. We have been working on methods to evaluate how these algorithms perform using prior results collected from PD patients using a closed-loop system to treat rest tremor. In this paper we propose metrics for closed loop algorithm evaluation based upon results from patient trials using wearable sensors and several example control algorithms.
\section{Methods}\label{sec:Methods}
We propose a metric that captures the trade-off between power efficiency and symptom suppression. The purpose of this metric is to compare the effectiveness of different dynamic systems across various patients. It bases this comparison by normalizing the closed-loop algorithm performance to the patient's clinical open loop tremor suppression and stimulation power used.

We first calculate an ``average tremor" value for both open-loop and closed-loop trials by summing the total band power in the gyrometer tremor band (4-8Hz) and dividing by the total tremor band power while the patient received no stimulation. From this, the additional tremor the patient experiences due to using the closed-loop DBS system is determined by subtraction as shown:
\begin{equation}
\label{tremor_inc}
\Delta Trem  = \cfrac{BP_{CL}}{BP_{No}} - \cfrac{BP_{OL}}{BP_{No}}
\end{equation}
where $BP$ is the average tremor band power while the patient was receiving closed-loop ($BP_{CL}$), open-loop($BP_{OL}$), or no stimulation($BP_{No}$).

In order to compare how an algorithm's delivered stimulation would impact the device's battery life, we can compare the stimulation power between the open-loop and closed-loop systems with the assumption that the electrode-tissue impedance is constant for the duration of the evaluation. With this assumption, the stimulation amplitude squared ($V_{CL}^2$) can be used to determine a normalized stimulation power value during the closed-loop trial. By then summing the instantaneous power across the trial and dividing by the total closed-loop trial duration ($T$), a normalized stimulation power average can be obtained. The algorithm's power reduction is then determined by dividing the average closed-loop stimulation power by the average open-loop stimulation power ($V_{OL}^2$) as shown: 
\begin{equation}
\label{pwr_red}
\Delta Pwr = 1 - \cfrac{\sum{V_{CL}\left(t\right)^2}}{T*V_{OL}^2}
\end{equation}

These two factors can be combined to obtain a single value that represents the trade-off that an algorithm brings to a patient between stimulation power and symptoms. A system performance metric that can be used to examine the trade-off between stimulation power and tremor mitigation is defined by:
\begin{equation}
\label{eq:perfmet}
M = \cfrac{\Delta Pwr}{100* \Delta Trem}
\end{equation}
where M represents the percent power reduction the closed-loop algorithm delivers for every one percent increase of tremor. An ideal closed-loop algorithm would save power at the cost of very little additional tremor for the patient, resulting in a high metric value. A poor algorithm on the other hand, my provide very little power savings (making the PWR term close to 0), or leave the patient with too much residual tremor (making the Trem term large). Such a poor performance system would result in a low metric score. The intuition behind this metric is that it captures the efficiency of a given algorithm at saving power balanced against additional tremor that the algorithm is not able to treat.

It is also important to consider how the metric performs in edge cases, in particular what would happen in the case of simply turning the stimulator on or off for the duration of the trial. In an ideal patient who has no tremor while the device is on ($BP_{OL} = 0$), turning the stimulator off ($V_{CL} = 0$) would result 100\% power savings but an 100\% increase in tremor back to the ``no-stimulation" level ($BP_{CL} = BP_{No}$). This would result in a trade-off value of 1\% power-for-tremor. On the other hand, simply leaving the stimulator on for the duration of the trial would result 0\% increase in tremor (since $BP_{CL} = BP_{OL}$) and 0\% power savings (since $V_{CL} = V_{OL}$), leading to an undefined result. This is expected since the metric is designed to compare algorithms to the open-loop performance in a normalized fashion. If the tested algorithm is equivalent to open-loop, then there will be no expected performance difference to compare.

\subsection{Controller Algorithms}
\begin{figure}[h]
\begin{center}
\includegraphics[height=2.5in]{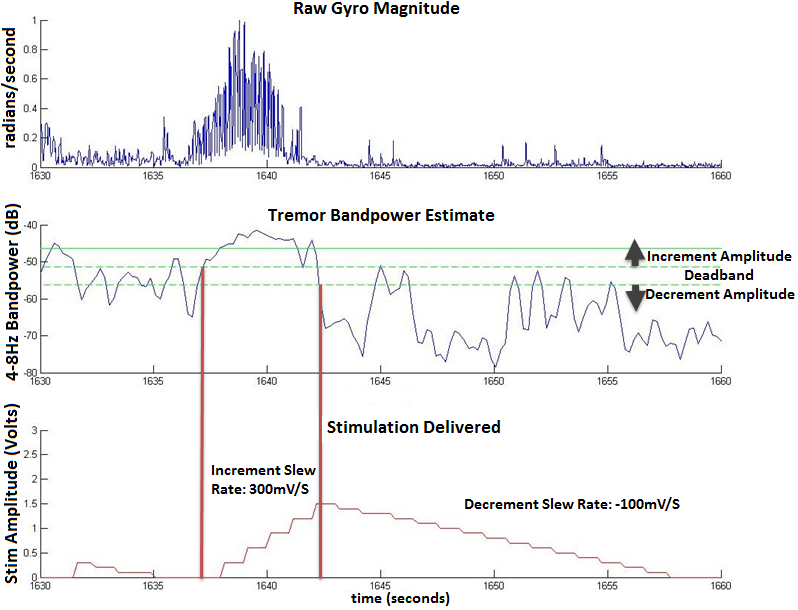}
\includegraphics[height=2.5in]{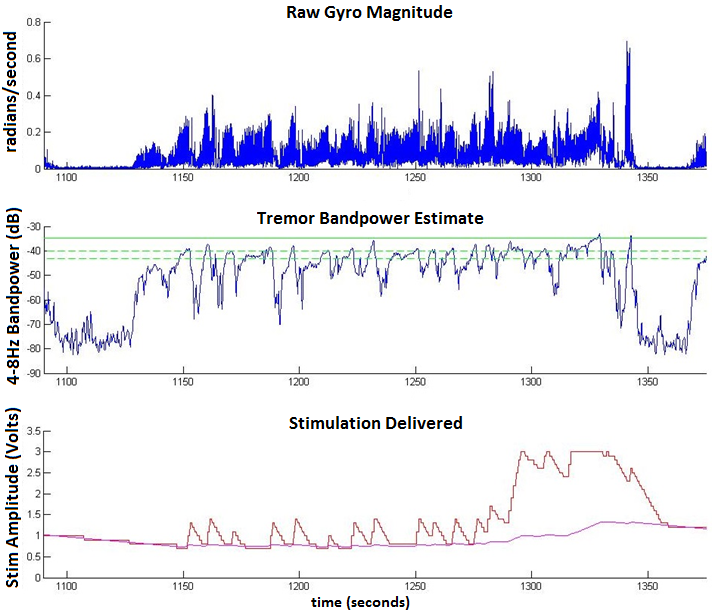}
\caption[Control Overview]{\textbf{Test Control Algorithms Overview} \footnotesize{Close-ups of the system responses when using each tested control algorithm: raw gyro on top (deg/s), tremor estimate (dB) in middle with calibration (solid horizontal) and thresholds (dashed horizontals), and output stimulation amplitude on bottom. \textit{Left: Threshold-based System, Right: Baseline-Modulating System.} Note bottom plot of the baseline system shows both the slow-changing baseline level and the instantaneous stimulation delivered.}}
\vspace{-5mm}
\label{fig:algorithm_overview}
\end{center}
\end{figure}

We evaluated our metric based on extended results from the closed-loop experiments for PD rest tremor that we have already published \cite{mahsa2015feasibility}. For these experimental projects we used the Medtronic Activa PC+S deep brain stimulation and the Nexus-D system, which provides a real-time communication interface between a computer and the Activa PC+S. This allows a host application running on the desktop computer to log data from the implanted electrodes and update the stimulation parameters. The algorithms tested relied on calculating a tremor estimate derived from the spectral band power in the tremor band from gyroscopic data collected via a smartwatch. While PD rest tremor is typically considered to have a peak between 4-6Hz \cite{jankovic2008parkinson}, we sum the 4-8Hz bins  in order to collect the entire band power of the fundamental tremor frequency. This tremor estimate was then used by two simple control algorithms shown in Figure \ref{fig:algorithm_overview}.

The first control algorithm compares the band power of the tremor to two thresholds to determine what action the stimulator should take. When the tremor estimate is above the high threshold, the stimulation amplitude will be increased. Similarly, the stimulation amplitude is decreased when the tremor estimate is below the low threshold. The separation between the two thresholds constitutes a dead-band where there is no change to the stimulation parameters. For the second control algorithm, we have two feedback loops: one fast loop to mitigate tremor as quickly as possible, and a second much slower loop to adjust a baseline stimulation to slow the re-emergence of tremor.

\subsection{Experimental Trials}
The pre-operative selection criteria, surgical technique, and assessment of subjects have been previously described \cite{bronte2009stn}\cite{bronte2010clinical}. The experiments were performed at least 6 months after initial programming visit when the stimulation settings were considered clinically optimized. Before trials, all subjects withheld long- and short-acting dopaminergic medication for more than 24 and 12 hours respectively prior to testing. The stimulation therapy contacts determined by the clinician were used during testing. For all subjects the stimulation frequency was 140Hz and the pulse width was 60 microseconds. Table \ref{table:demographics} shows the age and disease duration for each subject tested with a control algorithm discussed in the results section. Note that the first two STNs are from the same patient on different sides of their body. Additional information about the clinical feasibility and tolerability of using this system with patients is available in a separate paper \cite{mahsa2015feasibility}. We tested our example closed-loop algorithms with these patients and monitored their symptom's response when closed-loop stimulation was delivered to the STN contralateral to the tested limb. We then selected several interesting cases from a diverse set of trials to use as example cases to develop and evaluate the proposed metric.

\begin{table}[h!]
\begin{center}
\begin{tabular}{| p{.6cm} |p{1.6cm} | p{.6cm} | p{1.5cm} | p{1.8cm} | p{1.2cm} | p{1.5cm} | p{1.5cm} | p{2cm} |}
\hline\footnotesize{\textbf{STN ID}} & \footnotesize{\textbf{Controller}} & \footnotesize{\textbf{Age}} & \footnotesize{\textbf{Disease Duration}} & \footnotesize{\textbf{Implanted Time}} & \footnotesize{\textbf{Tested Limb}} & \footnotesize{\textbf{UPDRS Off DBS}} & \footnotesize{\textbf{UPDRS On DBS}} & \footnotesize{\textbf{Clinical Amplitude}} \\ \hline
1 & Threshold & 69 & 12 years & 10 months  & RHand & 15 & 2 & 2.5 Volts \\ \hline
2 & Threshold & 69 & 12 years & 10 months  & LHand & 15 & 2 & 2.5 Volts \\ \hline
3 & Baseline & 63 & 2 years  & 18 months & LFoot & 21 & 1 & 3 Volts \\ \hline
4 & Baseline & 72 & 9 years  & 7 months  & RHand & 15 & 6 & 2.6 Volts\\ \hline
\end{tabular}
\caption[Subject Demographics]{\textbf{Subject Patient Demographics:} \footnotesize{STN ID refers to which experimental trial the subject participated in as described in the results section. Disease duration is when symptoms were first reported by the patient. Reported UPDRS scores for each patient are total lateral UPDRS III scores for the limb tested while off medication.}}
\label{table:demographics}
\vspace{-5mm}
\end{center}
\end{table}

\begin{figure}[h!]
\begin{center}
\includegraphics[width=6.4in]{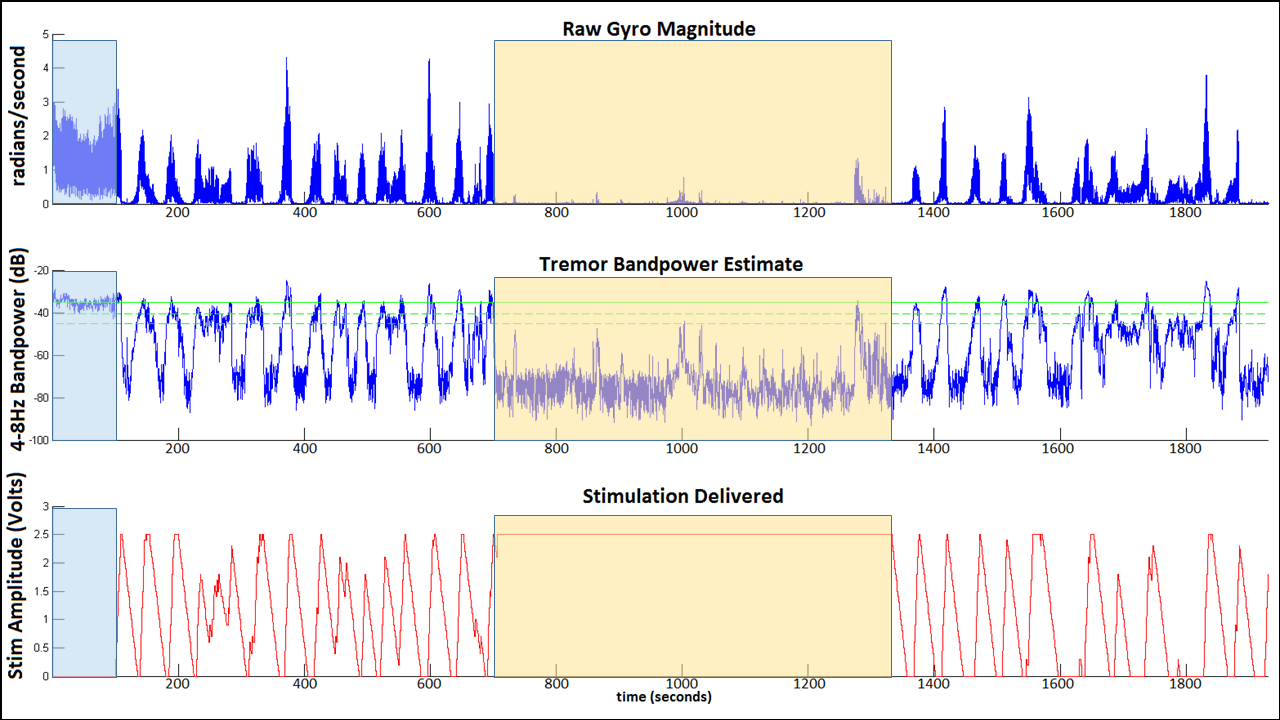}
\includegraphics[width=6.4in]{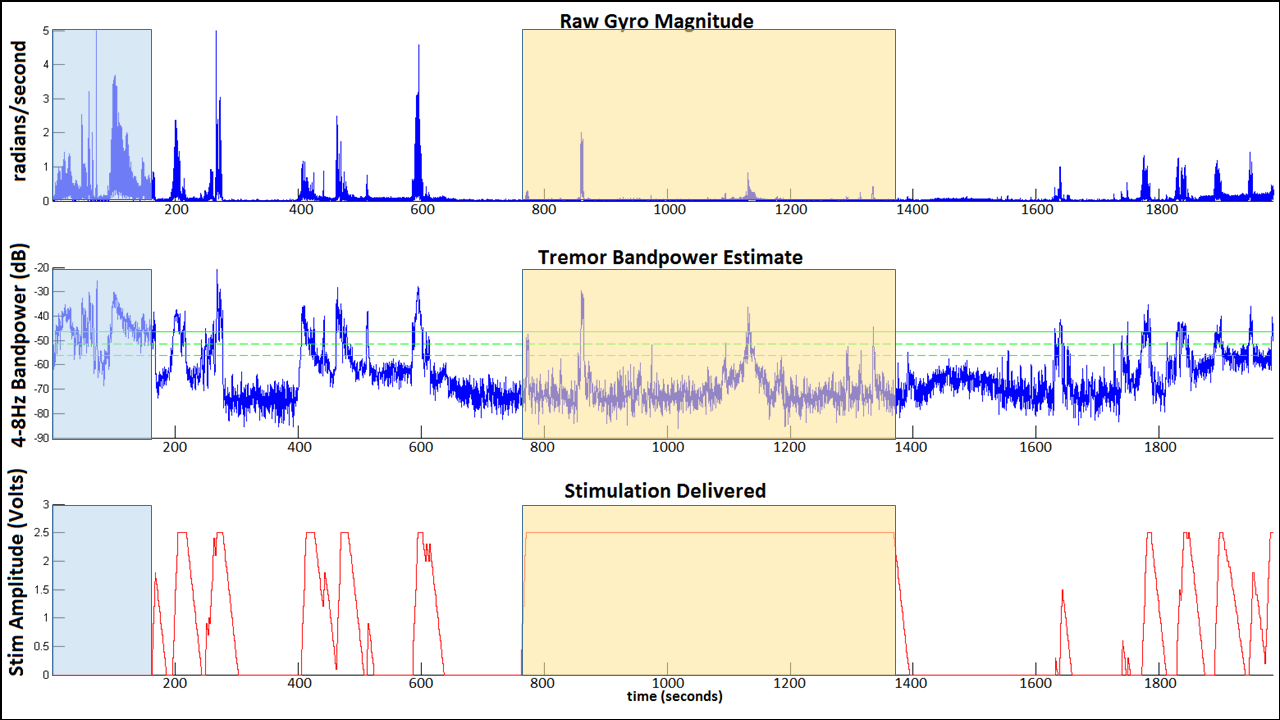}

\caption[Threshold Trials]{\footnotesize{\textbf{Threshold Trial Data:} plots of the tremor and stimulation data collected while running the threshold system. Blue overlay indicate periods with no stimulation, yellow periods with clinical open-loop stimulation. During closed-loop, stimulation amplitude was ramped up at 0.3 volts/second and ramped down at 0.1 volts/second. \textit{Top-} Three axis gyroscope sensing magnitude (deg/s). \textit{Middle-} Tremor estimate in blue, calibrated no stimulation level in green, dead band thresholds in dashed green. \textit{Bottom-} Delivered stimulation amplitude in volts. \textbf{Top Trial - STN 1 (RHand):} No stimulation before t = 100s; closed-loop between 100s and 706s; open loop between 706s and 1332s; closed-loop after 1332s. \textbf{Bottom Trial - STN 2(LHand):} No stimulation before t = 161s; closed-loop between 161s and 762s; open loop between 762s and 1390s; closed-loop after 1390s.}}
\vspace{-5mm}
\label{fig:threshold_results}
\end{center}
\end{figure} 

\begin{figure}[h!]
\begin{center}
\includegraphics[width=6.4in]{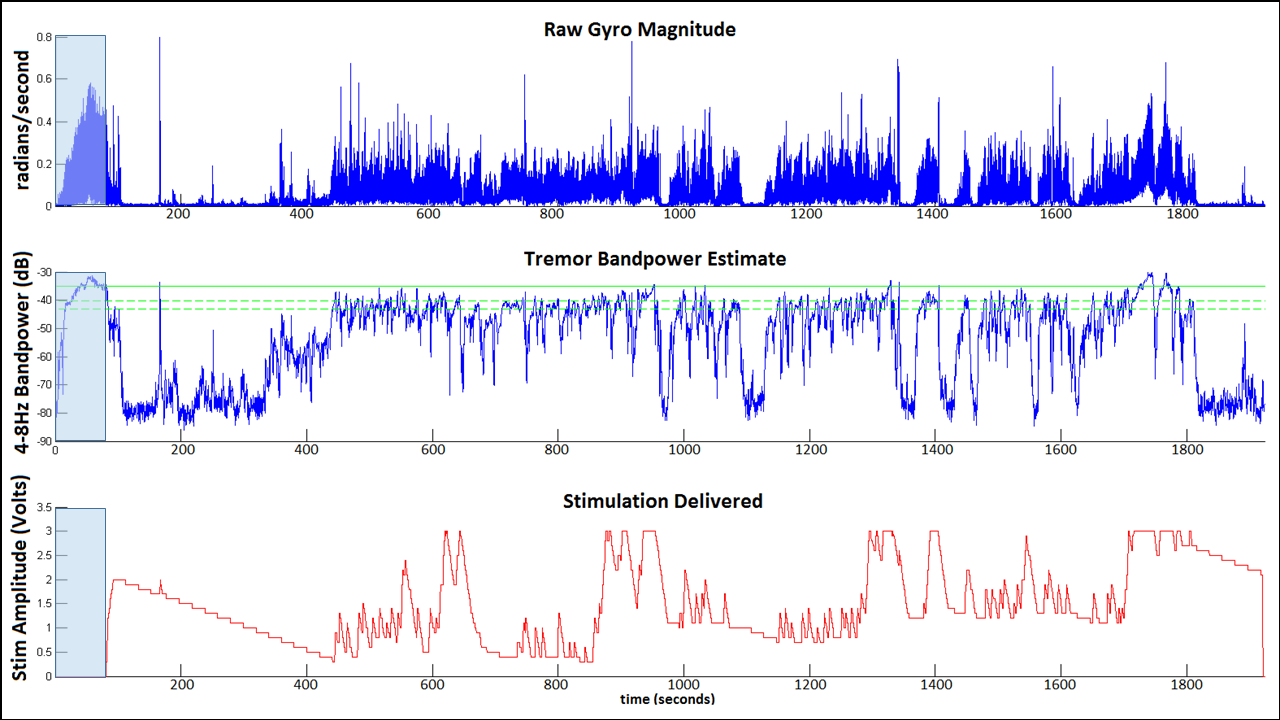}
\includegraphics[width=6.4in]{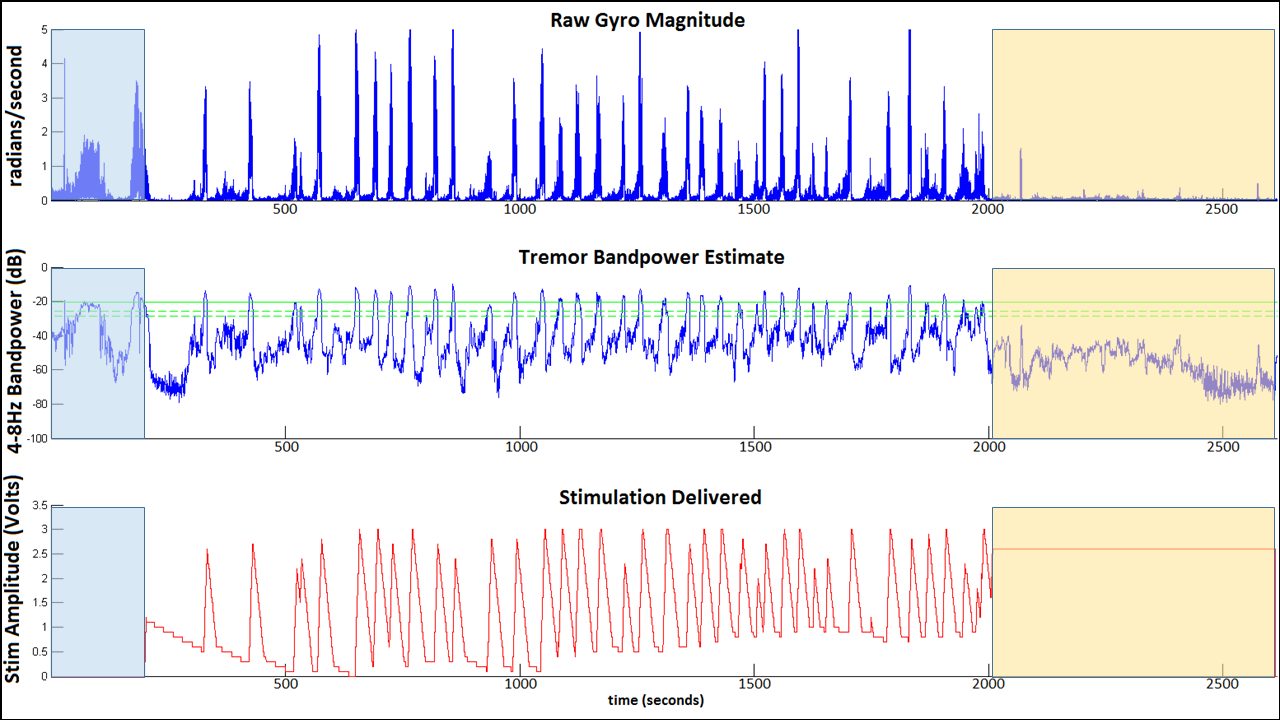}
\caption[Baseline Trials]{\footnotesize{\textbf{Baseline Trial Data:} plots of the tremor and stimulation data collected while running the baseline system. Blue overlay indicate periods with no stimulation, yellow periods with clinical open-loop stimulation. During closed-loop, stimulation amplitude slew rates were 0.3 volts per second upward and 0.1 volts per second downward. The baseline controller incremented 100mV every 4 seconds of tremor and decrementing by 100mV every 20 seconds without tremor. \textit{Top-} Three axis gyroscope sensing magnitude (deg/s). \textit{Middle-} Tremor estimate in blue, calibrated no stimulation level in green, dead band thresholds in dashed green. \textit{Bottom-} Delivered stimulation amplitude in volts. \textbf{Top Trial - STN 3 (LFoot):} No stimulation before t = 80s; closed-loop after 80s. \textbf{Bottom Trial - STN 4 (RHand):} No stimulation before t = 200s; closed-loop stimulation between 200s and 2011s; open loop stimulation after 2011s. }}
\vspace{-5mm}
\label{fig:baseline_results}
\end{center}
\end{figure}

The first two STNs come from the same patient when we tested the threshold based algorithm on each of their limbs. While the calibrated thresholds were unique to each side, the ramping and clinical stimulation parameters for both sides were the same. However, each limb responded quite dramatically different to the thresholding algorithm shown in Figure \ref{fig:threshold_results}. The right hand's tremor response to the threshold-based closed-loop system oscillated between treated and untreated states. The stimulation would rapidly bring the tremor to zero, which the system would then respond to by lowering the stimulation amplitude. Tremor often would not resume until some variable time after the stimulation amplitude was zero volts, upon which it would then re-emerge to untreated levels very quickly.In contrast to the right hand, the left hand's tremor was far more intermittent. This meant that after mitigating the tremor, it could often be minutes until the tremor re-emerged. While the open-loop system caused only one instance of tremor compared to the 5 in each closed-loop trial, the bursts of tremor were quickly mitigated for extended periods of time when compared to the our previous trial.

The moving-baseline algorithm was tested on two different patients (STN 3 and STN 4) and are shown in Figure \ref{fig:baseline_results}. For STN 3, The baseline was continuously adjusted as tremor was estimated and the fast loop was used whenever the estimate was above the threshold. The response of the patient's tremor to the system was far less oscillatory and there were no longer the distinctive treated/untreated states that were present in the previous trials. The tremor estimate was also consistently less during the closed-loop paradigm then in the calibration period. While the baseline seemed to reduce the oscillatory responses in the first baseline patient, the results from the second subject show a return to large oscillations between treated and untreated states. One difference between the two patients is the speed with which this patient responds to stimulation to mitigate tremor. In this trial tremor appeared, disappeared and reappeared quite quickly, with the stimulation amplitude never remaining at the clinical value for more than several seconds at a time. Additionally, many of the peak tremor estimations are far above the tremor seen during the calibration period. It also seemed as though the moving baseline stayed quite low for the duration of the trial but was slowly trending upwards. In control terms, it may be that the gain of the fast-loop was simply too high and drove the system to oscillate and the baseline loop wasn't able to respond on the experiment's timescale.
\section{Results}
Using our metric, the quantitative performance of the four trials are presented in a single table shown in table \ref{table:results}. The table compares open-loop to closed-loop symptom measures by normalizing to the no stimulation tremor as discussed in the methods section. Separate metric terms are shown before calculating the final proposed metric. Note that in all cases there is a qualitative decrease in average stimulation amplitude at a cost of an increase in tremor when comparing closed-loop to open-loop performance. For these example closed-loop algorithms, the stimulation power reduction ranged from 56\% to 82\%, while the tremor difference between closed-loop and open-loop ranged from 11\% to 94\% of the tremor while receiving no stimulation. The metric column for the trials illustrates the variation in system performance a single algorithm can have across patients or limbs.  While not used by our metric, the ``Tremor Time \%" term is also presented that describes the portion of the trial where the tremor was above the increment threshold. This indicates the proportion of the trial that the tremor was at an undesirable level.

\begin{table}[h!]
\begin{center}
\begin{tabular}{| p{1.15cm} || p{.8cm} | p{1.5cm} | p{1.3cm} || p{.8cm} | p{1.5cm} | p{1.3cm} || p{1.2cm} | p{1.1cm} | p{1.05cm} ||}
\hline
\multicolumn{1}{|c||}{ } & \multicolumn{3}{|c||}{Open-Loop Results} & \multicolumn{3}{|c||}{Closed-Loop Results} & \multicolumn{3}{|c|}{\textbf{CLDBS Metrics}}
\\ \hline
Tested STN & Stim Amp $V_{OL}$ & Tremor $\left(\cfrac{BP_{OL}}{BP_{No}}\right)$ & Tremor Time\% & Max Stim Amp & Tremor $\left(\cfrac{BP_{CL}}{BP_{No}}\right)$ & Tremor Time\% & Power Red. $\Delta Pwr$ & Tremor Inc. $\Delta Trem$ & \textbf{Perf. Met. ($M$)} \\ \hline \hline
1 (Th) & 2.5V & 0.45\% & 0.35\% & 2.5V & 27.2\% & 19.8\% & 72.2\% & 26.7\% & \textbf{2.7\%} \\ \hline
2 (Th) & 2.5V & 5.31\% & 2.75\% & 2.5V & 16.8\% & 12.7\% & 82.6\% & 11.5\% & \textbf{7.2\%} \\ \hline
3 (BL) & 3V & 0\% & 0\% & 3V & 19.4\% & 13.0\% & 56.9\% & 19.4\% & \textbf{2.9\%} \\ \hline
4 (BL) & 2.6V & 0.28\% & 0\% & 3V & 95.1\% & 16.2\% & 66.4\% & 94.8\% & \textbf{0.7\%} \\ \hline
\end{tabular}
\caption[Results Overview]{\textbf{Closed-Loop Performance Metrics:} \footnotesize{``Stim Amp" values are the constant stimulation amplitude during open-loop trials and are the maximum allowable amplitude for closed-loop trials. ``Average Tremor" values are dividing the average tremor band power in the loop mode by the average tremor band power with no stim. ``Tremor Time \%" is the portion of trial where the tremor band power is above the threshold to increase stim. ``Power Red." is the reduction in stim power across the closed-loop trial when compared to the open-loop stim trial. ``Tremor Diff" is the difference between open-loop and closed-loop average tremor values. The final closed-loop performance metric is based on equation \ref*{eq:perfmet}. Note: STN 3 open-loop tremor was not collected as part of the trial but the patient was clinically evaluated to have no tremor with DBS on.}}
\label{table:results}
\vspace{-5mm}
\end{center}
\end{table}
\section{Discussion}\label{sec:Discussion}
With this metric we can present algorithm efficiency directly to the clinicians evaluating closed-loop systems for their patient for easy comparison of control algorithms. By combining the performance scoring into a single value that indicates the tremor/stimulation trade-off, we also make it easier for future developers of closed-loop systems to benchmark their algorithms. Higher scores indicate that a system is capable of more efficient treatment of the patient's symptoms. Additionally, by normalizing the parameters used to the untreated tremor level and open-loop stimulation power, we are able to evaluate how a given algorithm performs broadly across patients whose symptoms and open-loop settings are different.

For the systems we evaluated, the thresholding algorithm when tested on the STN 2 patient's left hand produced the best results, where the proposed metric result shows that the system provided a 7.2\% decrease in power for every 1\% increase in tremor. Given how much power was saved with only modest tremor gains, a patient may find such a system to be an acceptable trade-off. On the other side of the spectrum, the baseline algorithm performed the worst during the STN 4 patient trial. In this case, the patient's response to the closed-loop system was highly oscillatory with peak tremor higher than the no stimulation case. The closed-loop system effectively compressed 95\% of the patient's no stimulation tremor into just 16\% of the closed-loop trial duration. For this patient, the proposed metric reveals that the system only provided a 0.7\% decrease in power per percentage point of increased power. Given the amount of tremor that the patient experienced during the trial, it is a given that such a system would be not suitable for this patient in it's current form. However, if properly tuned to the patient's specific dynamics, it is possible that such a system could perform much better in the future. The metric proposed by this paper gives us the tool to evaluate these future tuned systems and compare them to the prior tested algorithms shown in this paper.  

It will be important for future designers of the next generation of neuromodulation platforms to consider how their algorithms will be evaluated and engineer their systems to support the tests. In the future there may also be the need for additional terms in the metric as we learn more about engineering closed-loop neurostimulation systems. This may include power consumption from non-stimulation components, such as telemetry, or additional symptom metrics such as the ``tremor time" field in table \ref{table:results}. However, the metric is designed such that it is easily extensible to include additional terms and weighting by adding terms as desired, while also providing a tool that can be used to evaluate research algorithms with the current generation of hardware.

\section{Conclusion}\label{sec:Conclusion}
Given how rapidly the field of closed-loop neuromodulation research is expanding and how patient-specific the preliminary results have been so far, this work is a first step at developing methods for evaluating dynamic neuromodulation algorithms on a per-patient basis in a way that we can enable iterative development for the next generation of control algorithms. Defining the means for algorithm evaluation is also important for the device designers of the next generation of implantable systems so that they can include the tools required to do so. In this paper, we have presented a metric for closed-loop DBS algorithm evaluation that takes into consideration both normalized stimulation power savings and symptom treatment changes. This work builds upon experimental data taken from example closed-loop DBS systems that perform dynamic parameter adjustment using wearable inertial sensors. While the results of these four experimental trials vary, they provide an interesting lens to examine how to evaluate and compare different closed-loop algorithms and their interactions with patients.

\section*{Acknowledgment}
The University of Washington team is supported by a donation from Medtronic and by Award Number EEC-1028725 from the National Science Foundation for the Center for Sensorimotor Neural Engineering. The content is solely the responsibility of the authors and does not necessarily represent the official views of the National Science Foundation or Medtronic. The Stanford team is supported by the Michael J Fox Foundation for Parkinson's Research. 

\footnotesize
\bibliographystyle{unsrt}
\bibliography{Bibleography}
\end{document}